\documentclass[journal]{IEEEtran}

\IEEEoverridecommandlockouts

\usepackage{graphicx,color}
\usepackage{amsthm,amssymb,amsmath}
\usepackage{algorithm}
\usepackage{multirow}
\usepackage{algpseudocode}
\usepackage{footmisc}

\theoremstyle{definition}

\newcommand{\ie}{i.e.}
\newcommand{\eg}{e.g.}
\newcommand{\mc}[1]{\mathcal{#1}}

\title{A Novel Communication Cost Aware Load Balancing in Content Delivery Networks\\using Honeybee Algorithm}

\author{
	Hamid~Ghasemi, Mahdi~Jafari~Siavoshani, Saeed~Hadadan
\thanks{H.~Ghasemi, M. Jafari Siavoshani, and S.~Hadadan are with the Department of Computer Engineering, Sharif University of Technology, Tehran, Iran (emails: hghasemi@ce.sharif.edu, mjafari@sharif.edu, saeedhd@ce.sharif.edu).}
}

\begin{document}
\maketitle

%---------------------------------------------------------------------
\begin{abstract}
Modern web services rely on Content Delivery Networks (CDNs) to efficiently deliver contents to end users. In order to minimize the experienced communication cost, it is necessary to send the end user's requests to the nearest servers. 
However, it is shown that this naive method causes some servers to get overloaded. Similarly, when distributing the requests to avoid overloading, the communication cost increases. This is a well-known trade-off between communication cost and load balancing in CDNs.

In this work, by introducing a new meta-heuristic algorithm, we try to optimize this trade-off, that is, to have less-loaded servers at lower experienced communication cost. This trade-off is even better managed when we optimize the way servers update their information of each others' load. The proposed scheme, which is based on Honeybee algorithm, is an implementation of bees algorithm which is known for solving continuous optimization problems. Our proposed version for CDNs is a combination of a request redirecting method and a server information update algorithm. 

To evaluate the suggested method in a large-scale network, we leveraged our newly developed CDN simulator which takes into account all the important network parameters in the scope of our problem. The simulation results show that our proposed scheme achieves a better trade-off between the communication cost and load balancing in CDNs, compared to previously proposed schemes.
\end{abstract}

\vspace{7pt}
\begin{IEEEkeywords}
Content delivery networks, distributed load balancing, communication cost, Bees algorithm, Query cost.
\end{IEEEkeywords}

%---------------------------------------------------------------------
\section{Introduction}\label{sec:Introduction}
By increasing Internet utilization and emerging of new Internet-dependent applications, some popular websites and application servers generate high amounts of network traffic. As predicted in \cite{stocker2017growing}, until 2021, 71\% of web traffic will be served by Content Delivery Networks (CDNs). This number used to be about 52\% in 2016. A CDN is a group of network devices grouped together aiming at more effective delivery of content to end-users \cite{douglis2001guest}. In such a system, hundreds of servers are distributed over the world, each of them responsible for responding to geographically nearby users. However, it is obvious that in a CDN, a single server cannot hold all the available data, so having a suitable content distribution algorithm among all the available servers is necessary. There are some challenging problems and key features that have a direct effect on a CDN's performance such as content replication techniques, cache management, load balancing, etc. \cite{2007taxonomy}.

Mapping each request to the most appropriate server has a vital role in a CDN's performance. In \cite{doc_tradeoff} and \cite{ourselves}, the authors have introduced two important metrics to determine the most suitable server to redirect each request to, namely, \emph{communication cost} and \emph{servers' load} (that can be translated to \emph{response time}). As it is shown in \cite{doc_tradeoff}, redirecting requests to the nearest server leads to an excessively increased response time, while choosing the appropriate server without any distance considerations causes CDNs to experience high communication costs. In this situation, having an algorithm which determines a server considering both communication cost and the servers' load would have a significant effect on the performance of the whole system. 

In this paper, we have used the stochastic optimization problem defined in \cite{ourselves} as our main problem definition. Our proposed load balancing method, which is based on the Artificial Bees Colony (ABC) algorithm (derived from the natural behavior of honeybees foraging for food resources \cite{honeybee_main}), aims to provide an acceptable solution of the above-mentioned optimization problem. Our simulation results show that this algorithm achieves a better \emph{response time-communication cost} trade-off curve compared to the previous results.
Besides, as CDN providers have different business plans, they tend to optimize their services in favor of one or the other of these two metrics. So, it is essential to propose an algorithm that let the CDN provider be able to choose the importance of each parameter and move on the trade-off curve easily. 

Another influential factor on a CDN's performance is the way servers update their information about the queue length of other servers\footnote{In fact the queue length can be considered as a measure of servers' load.}. The more accurate this information is, the more effective and efficient the dispatching decisions are made. In \cite{ourselves}, the authors supposed that servers are ideally aware of each others' load. However, this assumption is not valid in the real world scenarios. 
%According to [\comment{reference}], 

In practice, there are two principal methods addressing the information update issue, namely, \emph{periodic} update and \emph{piggybacking} update mechanisms. In the periodic scheme, every server sends its queue status to all other servers at certain time periods. In contrast, piggybacking works differently. In this method, when server X sends a request to server Y, it sends its current load status back to the server X after responding to the request. Our proposed scheme utilizes a combination of periodic and piggybacking to update server's information about each other.

The rest of paper is organized as follows.
In Section~\ref{sec:RelatedWork}, we review some of the important previous related work on load balancing in CDNs. We also introduce the Honeybee algorithm (HBA) \cite{bees_alg} and some inherited versions of it. 
Section~\ref{sec:ProbDef}, formally defines our problem and introduce the important metrics related to the request redirecting scheme. Section~\ref{sec:ProposedAlg} explains our proposed algorithm in detail, which is evaluated in Section~\ref{sec:PerfEval} using a newly developed event-based simulation tool. Finally, Section~\ref{sec:Conclusion} concludes the paper.

%---------------------------------------------------------------------
\section{Related Work}\label{sec:RelatedWork}

Many work have been conducted on request routing in CDNs, considering various metrics and goals \cite{lb1,energy_aware,greening}. Generally, the proposed methods and algorithms can be divided into two major groups, \ie, \emph{adaptive} and \emph{non-adaptive} algorithms \cite{2007taxonomy}. In the former, the algorithm will decide on request routing in real-time, based on some defined parameters such as network crowdedness, servers' load, communication cost, etc., which are related to the current system conditions. In the latter, the request routing decisions are based on predefined variables and heuristics. The simplest example of a non-adaptive algorithm is the round robin (RR) request routing scheme.

It is evident that the adaptive algorithms are more complex and need to gather necessary state information from the network during execution. Consequently, it will impose some overheads on the network bandwidth. However, employing the real-time data will help the algorithm to overcome flash crowds and unpredicted network problems. So, devising an algorithm with low overhead and high tolerance of network issues is a desirable task.

The power of multiple choices \cite{p2c} is a classic paradigm for load balancing which is used in many practical implementations of request routing algorithms. Simplicity and achieving much better results in comparison to a simple randomized method are the main advantages of this paradigm. In \cite{next_neighbor} the authors have implemented an extension to this paradigm, called \emph{next-neighbor load sharing}. In this method, each server can receive the clients' requests and forward them to the most appropriate server. The best server is chosen between a random server and it's next neighbor. The next-neighbor load sharing method leads to better response time as the authors claimed.
In addition, Gardner and Balter have used the power of two choices concept to propose a new theoretical dispatching model \cite{balter_gardner}. In this algorithm, each request is copied in multiple servers' queues. This way, the requests would wait in multiple queues at the same time, resulting in the minimum waiting time across queues. The authors showed that copying each request in only two servers will reduce average response time noticeably.

Some other published algorithms like Joint Shortest Queue (JSQ) consider crowdedness of servers as the most critical decision-making parameter of redirecting. In a later work, Control-Law Load Balancing (CLB) algorithm \cite{control_law} improves JSQ's results. However, it still neglects the communication cost and assumes replica sets are close together.

Considering the communication cost along with the maximum load of servers (that translates to response time) as the two critical decision-making parameters affecting request dispatching is the main goal of \cite{doc_tradeoff} and \cite{ourselves}. These papers show that there is a trade-off between communication cost and response time in a distributed network. To handle the above-mentioned trade-off, \cite{ourselves} introduces three basic algorithms each of which leads to a different trade-off curve. Furthermore, it assumes that each server has a limited cache size, and therefore, it can respond to a specific group of requests. This limitation increases the complexity of the redirecting algorithms.

In \cite{cost_based}, authors improved the CLB algorithm by adding a new parameter $D_T$ to control communication cost. Although the existing trade-off is not mentioned in this paper, the simulation results somehow admit the existence of such trade-off. Its results show that more constraints on communication cost lead to worse response time. Despite \cite{ourselves}'s assumption about limited cache size, the authors of \cite{cost_based} consider every server is capable of responding to every request which is a simplified model of real CDNs.
%\comment{add energy aware load balancing, if it is necessary}

\subsection{Bees Algorithm}\label{subsec:bees}
Bees algorithm is a population-based search scheme, which is categorized in the class of meta heuristic algorithms. The process is a mimicry of foraging behavior of honeybees to look for the best solution to an optimization problem \cite{honeybee_main}. 

Bees in the same colony collaborate to find the best places of existing flower patches (\ie, food sources). A small fraction of the colony is responsible for finding new flower patches and foraging constantly. They randomly move in the surrounding environment and evaluate the profitability of the founded nectars. When returning to the hive, those bees that have found new beneficial flower patches start a special dance referred to as \emph{waggle dance}. It is observed that the dance duration is proportional to the efficacy of the founded patch. The longer the duration of the dance is, the more worker bees enlist to start exploiting the new patch. By this process, the worker bees broadly spread through the best patches. Thanks to evaluating new source foods, they can replace an existing patch with new profitable ones.

\begin{figure}
	\begin{center}
		\includegraphics[width=0.5\textwidth]{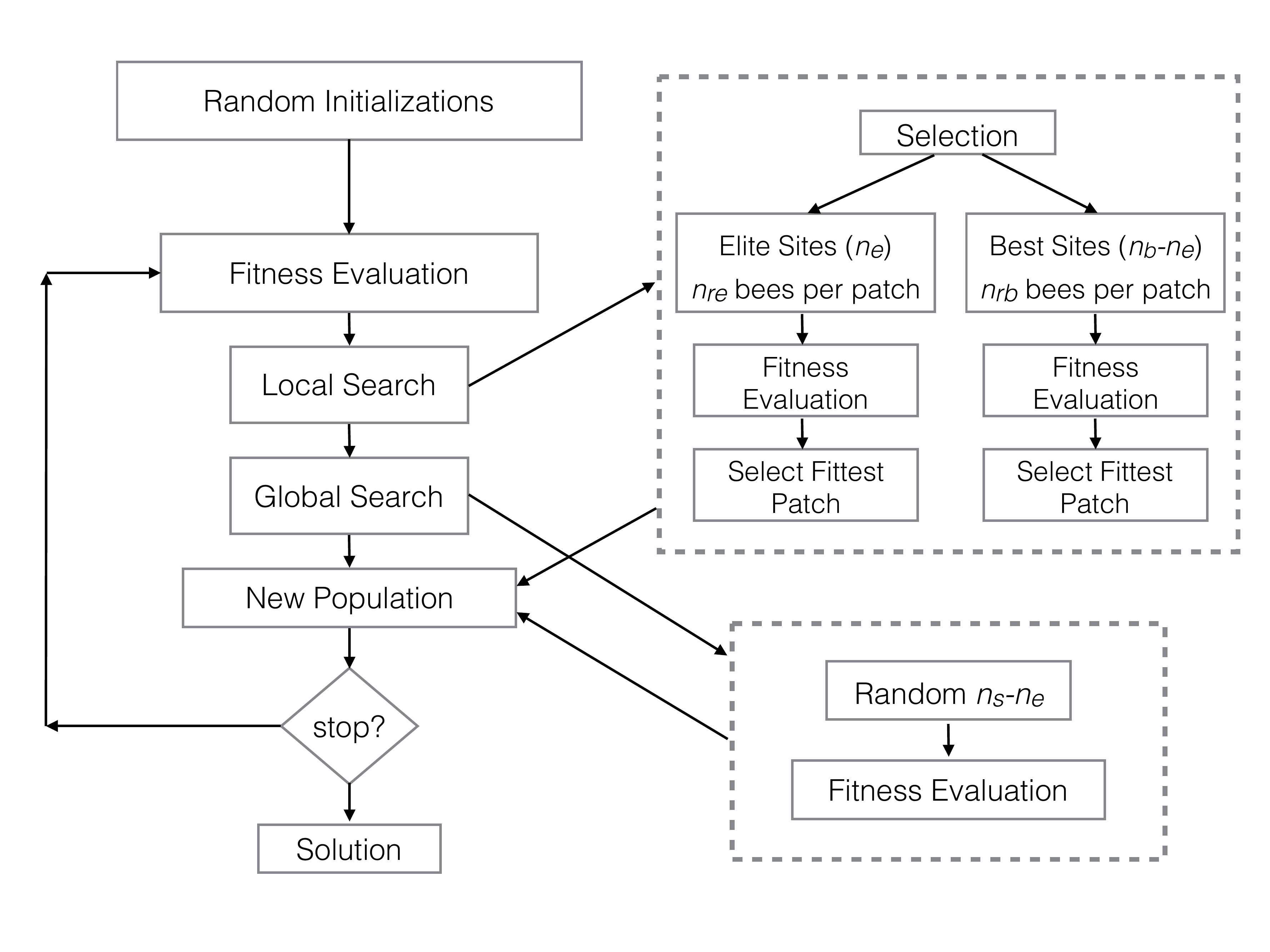}
	\end{center}
	\caption{Flowchart of basic bees algorithm \cite{bees_alg}.}\label{fig:honeybee}
\end{figure}

Honeybee algorithm (HBA) attempts to mimic honey bees' behavior in order to find the best results of an optimization problem. Fig~\ref{fig:honeybee} demonstrates the flowchart of basic bees algorithm.

Initially, $n_s$ scout bees are randomly spread through the search space with uniform probability. Then, the bees determine the fitness of the solutions in their landing place.
Next, the algorithm enters in a loop which contains four main steps.

The next procedure is called \emph{fitness evaluation}, where the scouts on the $n_b \le n_s$ fittest solutions start to recruit foragers by doing \emph{waggle dance}. These foragers are responsible for re-inspecting the best neighborhoods. $n_e \le n_b$ best neighborhoods are searched by $n_{re}$ foragers and the $n_b-n_e$ remaining best solutions are searched by  $n_{rb} \le n_{re}$ foragers.

In the \emph{local search} routine, the foragers locally rummage the flower patches of the scouts by which they were recruited. Any forager finding a better solution in a patch than what the scout has found in it will be designated as the new scout. At the end of each round, the fittest solution visited so-far is taken as the representative of the whole flower patch.

%Otherwise, when no forager finds a better solution, a neighborhood shrinking procedure steps in to shrink the patch size. This procedure causes the local exploration to be gradually focused on the fittest local solution. When repeating this local search for a pre-defined number of times leads to no improvement, the site is abandoned and the local maximum so far is considered the real local maximum. Then, a new scout is randomly generated and placed (site abandonment procedure).

In reality, there is a small portion of the bee colonies responsible for searching in the undiscovered parts of the search space which may be of high-fitness. This inspires the \emph{global search} procedure, a procedure that re-initializes $n_s-n_b$ scout bees to forage for new flower patches.

When each search cycle ends, again $n_s$ scouts are the new generation: the global search routine produces $n_s-n_b$ scouts in addition to the $n_b$ scouts produced by the local search procedure.  
In other words, we come to this equation for the total bee colony size: $n=n_e•n_{re}+(n_b-n_e)•n_{rb}+n_s$ which takes into account respectively, the elite sites foragers, remaining best sites foragers, and scouts.

It is obvious that the \emph{stopping criterion} depends on the problem's requirements. In some cases there is a pre-defined threshold which stops the algorithm after finding the first result which is better than the threshold. Also, the algorithm can run continuously and never stops.
 
%\comment{Is this paragraph needed?}The Honeybee Algorithm(HBA) formulation dates back to 2004 when Craig A Tovey at Georgia Tech with Sunil Nakrani at Oxford University wanted to devise a method for computer allocation to clients and web-hosting servers. In early 2005, Virtual Bee Algorithm (VBA) was taken advantage of by Xin-She Yang at Cambridge University. He used the VBA aiming to solve continuous optimization problems. In late 2005, Honey-bee mating optimization algorithm(HBMO) was developed by Afshar, Hadad and their coworkers. HBMO was later used for reservoir modeling and clustering.  As another extension, Artificial Bee Colony(ABC) was developed by Karabogo in Turkey, with the aim of numerical function optimization. 

In this paper, we introduce a new adaptive redirecting algorithm which emphasizes on reducing communication cost and response time simultaneously. Additionally, the CDN provider should be able to determine the importance of each parameter while routing the requests among existing nodes. In the next section, we will precisely define our problem and introduce parameters that are essential to be considered in a CDN.
%---------------------------------------------------------------------
\section{Problem Definition}\label{sec:ProbDef}

There are many parameters in a real-world CDN having impacts on its  performance. In the following, we introduce a reasonable model which contains many of these effective parameters. The proposed model is neither too simple to lose important details nor too complicated to be hard to analyze or simulate.

We model a CDN as a graph whose nodes represent replica servers (\ie, there exist $L$ servers), and edges represent direct network connections of the servers (the servers can be in the same or different data centers over the world). Our model mainly aims to deliver a set of $N$ files $\mc{W} \triangleq \{W_1 , W_2, \dots, W_N\}$ to the existing clients. We assume every file is of the same size $F$ bits, but their \emph{popularity} varies in the entire system that follows a specific probability distribution. Also, each server $i$ has a finite cache size where its content is denoted by $Z_i\subseteq\mc{W}$. We assume the servers' cache size are identical and equal to $M$ files, \ie, the cache size is equal to $MF$ bits or $|Z_i|=M$. 

\begin{figure}
	\begin{center}
		\includegraphics[width=0.45\textwidth]{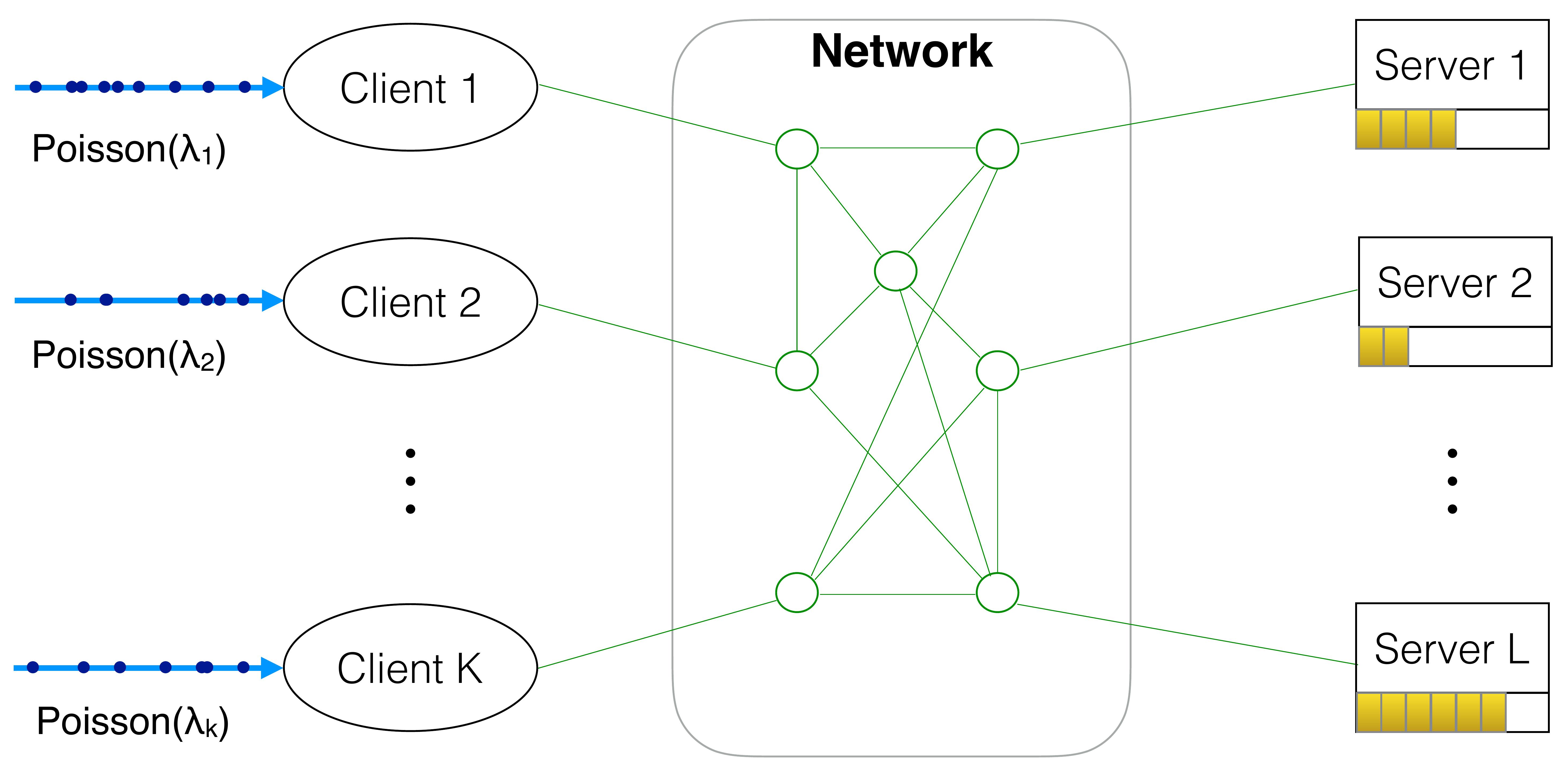}
	\end{center}
	\caption{A CDN with $K$ users and $L$ servers.}\label{fig:Model}
\end{figure}

Additional considerations on the proposed model are as follows (also see Fig.~\ref{fig:Model}):
\begin{itemize}
	\item
	There are $K$ users that send requests to their nearest server, based on Poisson distribution with parameter $\lambda_i$ for user $i\in [1:K]$.
	\item
	In each request, the probability of requesting file $W_i$ is $p_i$, where $\sum\limits_{i=1}^N p_i = 1$.
	This is called the popularity profile of files and based on previous observations (\eg, see \cite{Zipf1_99} and \cite{Zipf2_07}), we assume it follows the Zipf distribution, \ie,
    \begin{equation*}
    p_i=\frac{1/i^\beta}{\sum\limits_{j=1}^{N}1/j^\beta},\quad i=1,\dots,N,
    \end{equation*}
    where $\beta\ge 0$ is a parameter\footnote{Note that when $\beta = 0$, the Zipf distribution becomes the uniform distribution.}.
    \item
    We assume transferring a file from servers to clients, impose some communication cost to the CDN. To consider this cost, we define a cost matrix $\mathbf{C}\triangleq [c_{i,j}]_{i=1,j=1}^{i=K,j=L}$ which determines the cost of sending files from each server to each client.
	\item 
	All the queue states of CDN servers at time instance $t$ is represented by the vector $\mathbf{q}(t)\triangleq (q_1(t),\ldots, q_L(t))$.
    \item 
    The dispatching scheme at server $l$ is denoted by $\Psi_l$ that is a mapping defined as
    \begin{equation}
    \Psi_l \big( (i,j), \mathbf{C}, \mathbf{\tilde{q}}_l(t_{i,j}) \big) \mapsto \{1,\ldots,L\},
    \end{equation}
    where $(i,j)$ denotes the $j$th request of user $i$, $t_{i,j}$ represents the arrival time of this request, and $\mathbf{\tilde{q}}_l$ is the \emph{available estimation} of servers' load at the $l$th server. The whole dispatching scheme of CDN is denoted by $\mathbf{\Psi} = (\Psi_1,\ldots,\Psi_L)$.
\end{itemize}

As mentioned earlier, in this paper, we assume each server has access to part of the file library. However, the aggregate memory of the servers in the CDN stores the whole contents, that means we have $LM \ge N$. For simplicity, in the model above, there is no dynamic in the content placement strategy and putting the files in each server happens during the initialization phase of the CDN. However, adding dynamic of popularity distributing to the model is straightforward. 

\begin{figure}
	\begin{center}
		\includegraphics[width=0.45\textwidth]{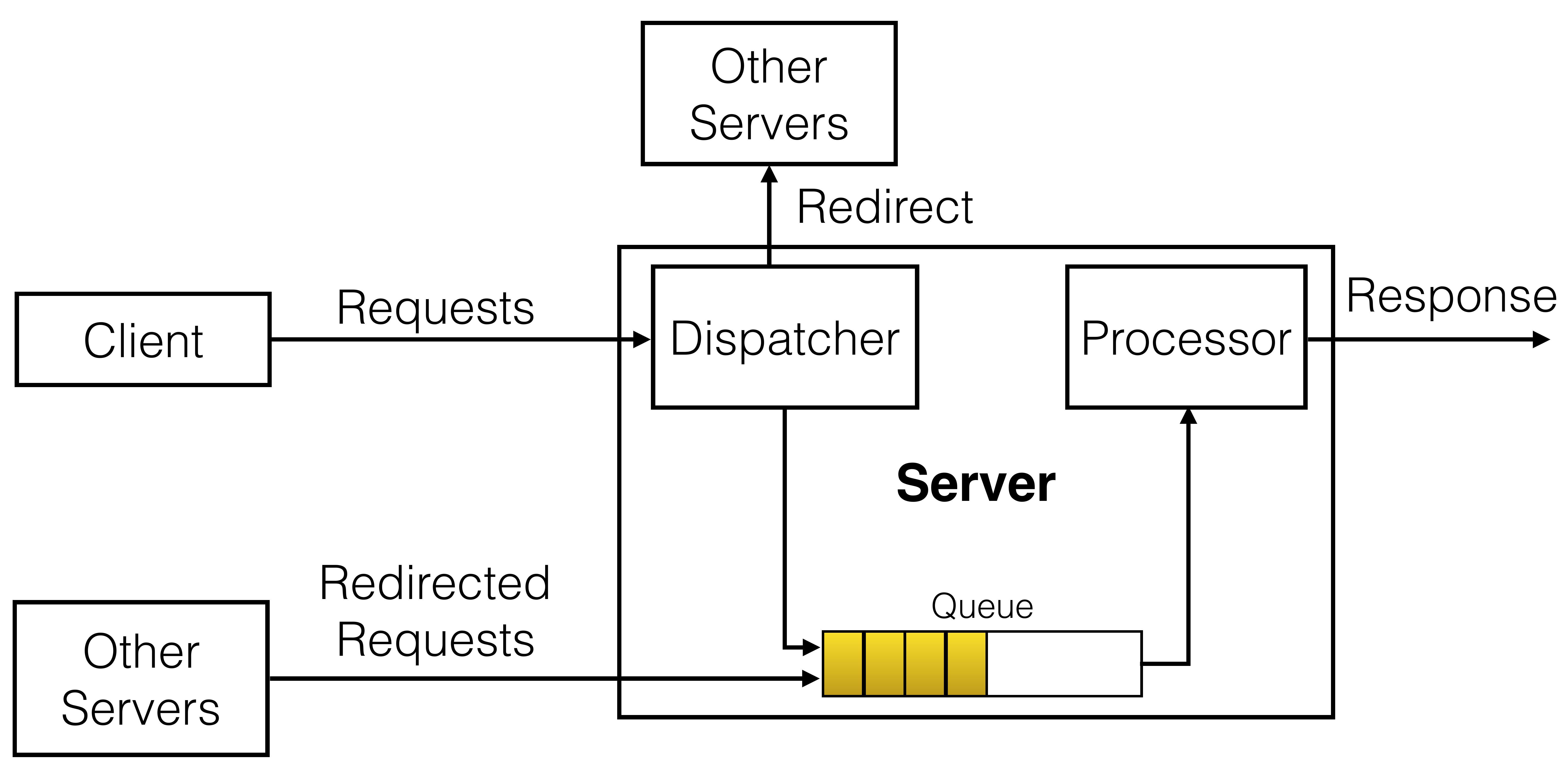}
	\end{center}
	\caption{In-server flow model. This figure illustrates what happens to the incoming requests to a given server.}\label{fig:queue_model}
\end{figure}

In this model, we assume when a client's request is received at a server, it is responsible for redirecting the request to the most appropriate server (including itself if it contains the requested file). In contrast, 
when a server receives an already redirected request from other replicas, it should serve that request without any further redirection. Fig.~\ref{fig:queue_model} shows the flow model of requests in each server.
Additionally, we assume the server's queue follows a first-in-first-out (FIFO) strategy and the process time for each request follows a particular distribution (\eg, exponential distribution).

After all, we are seeking a new redirecting algorithm which is able to decrease the communication cost along with the response time. To achieve this goal, it is necessary to have a well-defined definition on the communication cost and the average response time.

We define the average communication cost per file which is incurred to a CDN as follows
\begin{equation}
\bar{C} (\mathbf{\Psi}) \triangleq  \mathbb{E} \left[ \frac{1}{\sum_{i=1}^{K}{n_i}}\sum_{i=1}^{K}{\sum_{j=1}^{n_i}{c_{i,S_{i,j}}}} \right],
\end{equation} 
where $n_i$ is the number of requests sent by the client $i$ in a long time period $T$, and $S_{i,j}$ determines the server which is responsible for answering the $j$th request of the $i$th client, \ie, $S_{i,j} = \Psi_l \big( (i,j), \mathbf{C}, \mathbf{\tilde{q}}_l(t_{i,j}) \big)$.

Additionally, we can define the average response time as follows
\begin{equation}
\bar{D} (\mathbf{\Psi}) \triangleq  \mathbb{E} \left[ \frac{1}{\sum_{i=1}^{K}{n_i}}\sum_{i=1}^{K}{\sum_{j=1}^{n_i}{\tau_{i,j}}} \right],
\end{equation}
where $\tau_{i,j}$ is the duration from sending the $j$th request of the user $i$ until the receiving time of the first packet of the response by the client.

Having defined the above metrics, now we can formally capture the trade-off between communication cost and waiting time by the following optimization problem

\begin{equation}\label{eq:Trade-off_OptProblem}
\begin{aligned}
& & &	\min_{\mathbf{\Psi}} \Big[ a \bar{C}+(1-a)\bar{D} \Big] \\
&	\text{s.t.} & \\
& & & \text{Each request $(i,j)$ is routed by $\mathbf{\Psi}$ to a server that}\\
& & & \text{has cached the requested file,}
\end{aligned}
\end{equation}
where $a\in[0,1]$ is parameter that let us move on the trade-off curve. Notice that $a=0$ leads to the \emph{minimum response time strategy} and $a=1$ results in the \emph{minimum communication cost strategy}.

In the next section, we will introduce an extension of the Honeybee scheme to provide an acceptable solution to the above optimization problems. We will show through simulation evaluations that the proposed scheme achieves the best known trade-off curve for the above problem.

\section{Proposed Algorithm}\label{sec:ProposedAlg}

As it was mentioned in section~\ref{subsec:bees}, bees algorithm were initially introduced for solving optimization problems. We are going to use this algorithm to find the best servers to redirect the requests to. This section is divided into two parts. In the first part, we introduce our proposed fitness function. The Honeybee scheme utilizes this method to evaluate the results received by foragers. The second part is about the customization of Honeybee scheme. 
%In the following, we will explain the considerations and customizations in this algorithm.

\subsection{Fitness Function}
The proposed \emph{fitness function}, borrowed from \cite{ourselves}, is called \emph{Weighted Metrics Combination (WMC)} and described in Algorithm~\ref{alg:wmc}. It is shown in \cite{ourselves} that this algorithm produces the best ``response time-communication cost'' trade-off among the other proposed schemes, and our researches confirm that claim.

\begin{algorithm}[H]
	\caption{Weighted Metrics Combination (WMC) \cite{ourselves}. This algorithm is run in each server $l$.} \label{alg:wmc}
	\begin{algorithmic}[1]
		\Require $l$,$(i,j)$, $\alpha$, $\mathbf{C}$, $\mathbf{\tilde{q}}_l(t_{i,j})$
		\State $\Lambda \leftarrow \left\{k | k\in[1:L], d_{i,j}\in{Z}_k \right\}$ \label{algline:cache_cheking}
		\State Query $\{\tilde{q}_{l,k}(t_{i,j})\}_{k \in \Lambda}$
		\State $\beta_1 \leftarrow \sum_{k \in \Lambda} c_{i,k}$
		\State $\beta_2 \leftarrow \sum_{k \in \Lambda} \tilde{q}_{l,k}(t_{i,j})$
		\ForAll{server $k \in \Lambda$} 
		\State $\eta(k) \leftarrow \alpha \frac{c_{i,k}}{\beta_1} +(1-\alpha)\frac{\tilde{q}_{l,k}(t_{i,j})}{\beta_2}$
		\EndFor	
		\State Assign the request to $\underset{k \in \Lambda}{\arg\min} \ \eta(k)$
	\end{algorithmic}
\end{algorithm}

%Algorithm~\ref{alg:wmc} shows the pseudo code of this method. 
%In this algorithm we define $q(t_{i,j})$ as follows:
%\begin{equation}
%q(t_{i,j}) \triangleq (q_1(t_{i,j}), q_2(t_{i,j}) \dots q_L(t_{i,j})),
%\end{equation}
%where $q_l(t_{i,j})$ denotes the queue length of server $l$ on the arrival time of $j$'th request of user $i$.
In Algorithm~\ref{alg:wmc}, $\eta(k)$ is defined as the desirability of server $k$ for responding to the received request. A server with a minimum value of $\eta(k)$ would be the best server for redirecting the requests to. Note that in Line~\ref{algline:cache_cheking} of the code, we have excluded servers that do not have the requested file in their caches so that the chosen server certainly is able to respond to the request. To compare communication cost and crowdedness of servers, these two parameters are normalized by a simple scaling. In a nutshell, for each server $k$, the value of $\eta(k)$ is a weighted sum of the communication cost and current estimation of server $k$'s load (known at server $l$). The parameter $\alpha$ denotes the weight of each parameter. Obviously, when $\alpha = 1$, the communication cost is the only important parameter and when $\alpha = 0$, the load of servers determines the dispatching decision. 

\subsection{Our Customized Version of the Honeybee Scheme}
In the suggested request routing method, we are using an extended version of Bees algorithm which is customized for the defined problem. 

As we mentioned in section~\ref{subsec:bees}, bees are responsible for gathering information about the flower patches. We model each flower as a single server in our problem, and therefore, a group of geographically adjacent servers will form a flower patch\footnote{Note that the number of servers in each flower patch can be different.}. Each request received from the client side represents a bee in the Bees algorithm. 

According to the basic Bees algorithm, requests should be used for global and local search (in addition to responding them). To achieve this goal, we mandate each server, upon receiving a request, to list all the servers which contain the needed file (including itself if applicable) and redirect the request to one of these servers based on the following rules:

\begin{itemize}
	\item $\delta$ percent of the requests will be dispatched randomly among all servers (\ie, global search), and
	\item others will be redirected to the best server selected by the evaluation method stated in Algorithm~\ref{alg:wmc} (\ie, local search).
\end{itemize}

Using an evaluation method needs accurate information about the load of other servers. The more precise the information about other servers are, the better the result of load balancing will be. To achieve an optimum redirecting decision, we will use a combination of two updating methods. These methods are explained in the following.

Suppose server X redirects a request to server Y. After serving the request, server Y sends the load information of all servers in that patch to the server X (if they are not in the same patch). This way of sending information is inspired by the piggybacking update method which is a common way of sending back data using acknowledgment packets in computer networks \cite{tanenbaum}.

A significant difference in our implementation of Honeybee scheme from the basic algorithm is the periodic updates in our method. In the basic algorithm, the patches are completely passive and they can not exchange information. However, in a CDN, servers in the same patches would periodically exchange information about their load among themselves. It is obvious that the periodic updates among all servers in all patches add overheads to the network traffic and if the update periods decrease enough, it can even saturate all the network bandwidth. By using inter-patch updates, we benefit from the advantages of periodic update by adding a negligible amount of overhead to the network traffic. In Section~\ref{sec:PerfEval}, we will show the effect of inter-patch updates on the network in more details. 

%---------------------------------------------------------------------
\section{Performance Evaluation}\label{sec:PerfEval}
Since it is hard to derive the optimal solution of \eqref{eq:Trade-off_OptProblem}, in this section we evaluate the performance of proposed scheme  through preforming simulations using our own developed CDN simulator and compare the results to the results of existing schemes and algorithms. 

\subsection{Existing Testbeds}
According to their increasing deployment, CDNs have been under investigation on various testbeds developed in recent years \cite{CDNs,CDNSIM(Greek),testbed}. Among these testbeds, less than a handful are open-source or accessible to the public research community. These limited options can be categorized into two groups. First, software applications that simulate the problem, and second, test-purpose infrastructures that help researcher examine the solutions on a real, worldwide network. 
In the following, we discuss the characteristics of two simulation tools for CDNs, as well as a global testing infrastructure, referred to as PlanetLab. Then, we explain whether utilizing them was feasible for our problem or not. 

Our earliest attempt to emulate the problem was performed on NS2, but its excessive complexity prevented our experiments from being scalable. What is significant about NS2 is that it is a very realistic emulator, implementing a wide range of protocols currently in operation on Internet. This realisticity might seem to guarantee the reliability of the results. However, lots of these parameters in NS2 are not of noticeable impact on the results in our problem. Instead, many of these features appear to be time-consuming when running the simulator, especially for a large network. By comparing the results obtained from NS2 and our event-based simulator on small networks (that NS2 can handle the problem), it is evident that NS2 has a poor run-time performance, without considerable variation of the outcomes in our case.

CDNsim (developed in Aristotle university) \cite{CDNSIM(Greek)} is an open-source C++ program with a user-friendly GUI that covers a wide range of CDNs' various functionalities as declared in \cite{2007taxonomy}. Unfortunately, soon after trying to run CDNSim, it turned out that the simulator undergoes compatibility problems with the recent operating systems\footnote{After contact with the authors, we notified that it is not supported anymore.}.

Despite the former abolished simulator, a brand-new CDNsim (developed by CNPLab in NECLabs) decribed in \cite{CDNsim(German)1} and \cite{CDNsim(German)2}, which is written in Python, appears to be an alternative, publicly accessible option to experiment CDNs on it. It is a streaming-level simulator that enjoys a real-world topology of CDNs on the internet \cite{IRL}. However, being developed for the specific purpose of the producer(s) as well as ambiguous coding, made it entirely infeasible to customize for the purpose of our research.

In addition to the two simulation tools which were built as software applications,  PlanetLab \cite{PlanetLab} facilitates researching on CDNs but in a different fashion. PlanetLab is a global testbed service with physical equipment distributed around the world that serves diverse research fields in computer networks, including research on CDN technologies. Predictably, the responsible consortium has several requirements demanding equipment and personnel, in addition to a lingering formal procedure beyond our scope.

As none of the options above matched our problem needs, a specific-purpose simulator needed to be developed. It should have eliminated the parameters and factors of little impact in favor of performance. As a result, \emph{JCDNsim} has been developed from the scratch to fulfill the specific requirement of our problem plus offering an easily customizable platform for future exploits. Employing this futuristic perspective led to a cleanly coded Java-based simulator that can be simply perceived and extensively developed in the future \cite{JCDNsim}.

\subsection{JCDNsim Specifications}

JCDNsim is a brand-new, parallel, and discrete-event simulator written in Java. According to the several entities in JCDNsim, Java is a wise choice due to its Object Oriented structure. Also, when it comes to the increasing number of objects in the runtime, the memory management tools of Java appear critical. In this regard, our simulator leverages effective cooperation with Java garbage collector (GC). As a result, it is capable of increasing the number of servers, clients and requests, without worrisome about memory shortages. Additionally, runtime of simulating a large number of servers, clients, and requests, that used to be concerning us when using NS2, is drastically reduced. This is chiefly because of neglecting the assumptions that are not of noticeable impact on the problem.

As JCDNsim was intended to serve this research, it is completely adapted to our problem environment. Though, in case of expansion for further intentions, it is totally readable and modular as well as benefiting from Object Oriented Principles of low coupling and high cohesion.

Servers and clients in JCDNsim can interconnect in custom topologies. Each client is connected to a server to which sends its requests. The timing of requests generation are based on Poisson distribution: in exponential interarrival times, a random client will be selected that will send a request to its corresponding server for a file chosen based on the Zipf popularity distribution.

Data flow is designed as segments traversing the network through links that have predefined propagation delays and bandwidths. Files are distributed among the servers using Zipf distribution to simulate their popularities. The
ratio of the average \emph{input request rate} to the average \emph{service rate} is another important parameter of the simulator. Obviously, the value of this parameter should not be more than $1.0$, as it leads to an unstable system which gets overwhelmed with requests in the long run.

Due to various randomization sources in the problem (\eg, request arrivals, caches placement, etc.), in order to achieve a meaningful result, one has to perform many \emph{simulation runs} and average the results.
The number of required simulation runs depends on the precision needed for the comparison of results. 
In each run, the requests are regenerated using Poisson distribution, and the files are redistributed among the servers using Zipf distribution. Finally, the required statistics are gathered at the end of each run and the results are averaged at the end of all runs. All charts in the paper are obtained this way.

\subsection{Implemented Schemes}
JCDNsim implements both redirecting and update algorithms. As mentioned before, the implemented dispatching algorithm used in this study is WMC (see Algorithm~\ref{alg:wmc}). However, there are two other implemented redirecting algorithms proposed in \cite{ourselves}, namely, probabilistic scheme switching (PSS) and multiple choices scheme (MCS). Likewise, the implemented update algorithms are ideal, piggyback, and periodic. After all, in the proposed scheme, our customized version of Honeybee is also implemented, which uses a combination of piggyback and periodic for its update part and WMC for its redirecting part. Hereunder is the description of the aforementioned update algorithms. A summary is also provided in Table~\ref{table:algorithmms}.

\begin{itemize}
\item \textbf{Ideal}: An update algorithm that assumes each server ideally knows the instant queue length of all other servers in the network (this is in fact studied in \cite{ourselves}).

\item \textbf{Periodic}: It updates the servers at the end of certain time periods. At the end of each period, all servers send their load status to all other servers in the network. The duration of these periods is called \emph{update step}.

\item \textbf{Piggyback}: This update algorithm requires whenever each server serves a request, it should attach (piggyback) its load status to the response.
\end{itemize}

\begin{table}[ht]
	\caption{Implemented schemes.}
	\centering
	\begin{minipage}{7 cm}
	\begin{tabular}{c c c}
		\hline\hline
		Name &Type  & Parameters\\ [0.5ex] % inserts table %heading
		\hline 
		PSS	\footnote[1]{This scheme is implemented in the simulator, but not used in this study. \label{notused}}	&	Redirecting		&	$\zeta$ \\	
		MCS	\footref{notused}	&	Redirecting		&	$\Delta$ \\
		WMC		&	Redirecting		&	$\alpha$ \\
		\hline	
		Ideal & Update & 	$-$\\
		Periodic & Update & \emph{Update step} \\
		Piggyback & Update & $-$\\
		\hline
		\multirow{3}{*}{Proposed Sch.} & \multirow{3}{*}{Update and redirecting} & \emph{Update step} \\
		 & & \emph{Random search factor}\\
		  & & $\alpha$\\[1ex]
		\hline
	\end{tabular}
\end{minipage}
	\label{table:algorithmms}
\end{table}

\subsection{Configuration}
The configuration of JCDNsim can be easily changed by altering the parameters' values in the code or switching the Booleans in it. For the sake of comparing the proposed scheme with the existing algorithms, we set the parameters as follows. 

The default topology is formed by making each server connect to 3 servers with lower IDs and 3 servers with higher IDs. At the boundaries, we apply the same rule, but modulo the number of servers. Formally, the neighbors of server $s\in[0:n-1]$ is given by $\{j:  0\leq j< n, j \equiv_n s\pm i, i\in \{1,2,3\}\}$.
Finally, the degree is $7$ for each server, $6$ links to other servers and a single link to a client.

The number of servers and files are both set to $100$, and every server contains $10$ instances of different files (cache size). Basically, for every chart discussed here, the final chart is obtained with a set of $10^5$ Poisson generated requests plus a set of Zipf distributed files in the CDN. The number of simulation runs is set to $200$ to $10000$, based on the needed precision for the results. The \emph{input request rate} is set to $0.7$, which means that the rate of incoming requests is $0.7$ of the serving capacity of the whole system.

A summary of the configuration parameters for our further analysis are stated in Table~\ref{table:nonlin}.

\begin{table}[ht]
	\caption{Simulation configuration.}
	\centering
	\begin{tabular}{c c}
		\hline\hline
		Parameter & Value\\ [0.5ex] % inserts table %heading
		\hline 
		Servers&100 \\
		Clients&100 \\
		Files&100 \\
		Requests & $10^5$  \\ 
		Cache Size& 10\\
		Sites & 25  \\
		Bandwidth & 2 MB/s  \\ 
		Propagation Delay  & 0.1 ms  \\ 
		Request Size  & 0.5 KB  \\
		Simulation Runs & 200 to 10000\\[1ex]
		\hline
	\end{tabular}
	\label{table:nonlin}
\end{table}

\subsection{Determining Hyper-Parameters}
Before analyzing the trade-off between communication cost and load balancing (which is represented by ``average response time''), we need to determine the values of hyper-parameters. These hyper-parameters include: 
\begin{itemize}
\item \emph{update step} of the periodic algorithm,
\item \emph{update step} of the our proposed scheme, and
\item \emph{random search fraction} of the proposed scheme.
\end{itemize}

The optimum values for the above hyper-parameters help us compare the schemes at their best performance. By fixing constant values for the other parameters, we determine the optimum value of these hyper-parameters. 

\begin{figure}
	\begin{center}
		\includegraphics[width=0.4\textwidth]{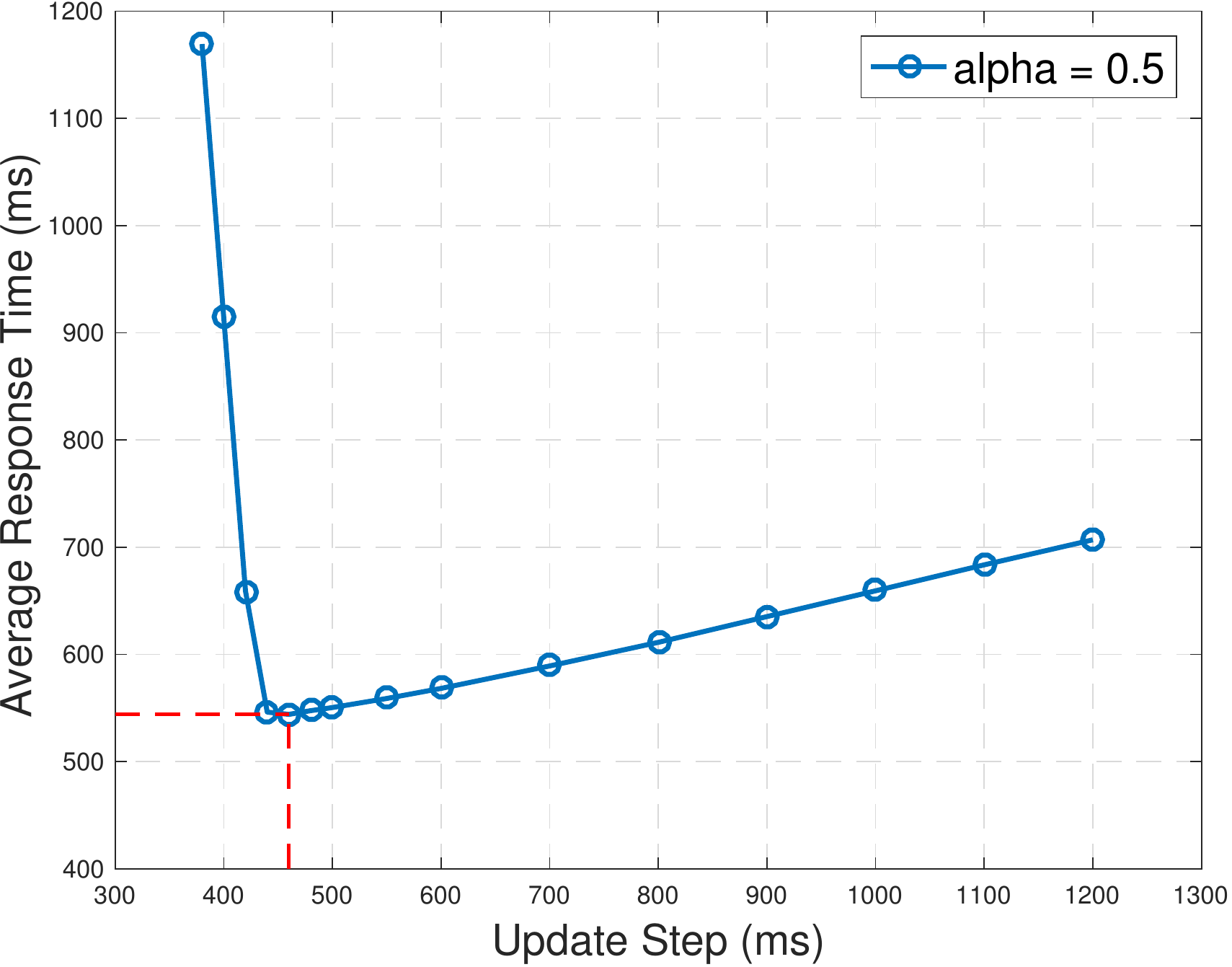}
	\end{center}
	\caption{Optimum value of update step in the periodic algorithm is $460$ ms.}\label{fig:P}
\end{figure}

First, we start by the update step of the periodic algorithm. Fig.~\ref{fig:P} illustrates how modifying this hyper-parameter can affect the average response time. It is also shown that the optimum value of the update step is $460$ ms. Expectedly, for the values more than $460$ ms, the curve shows lowering the update step leads to less response time, because the news from the other servers is updated more frequently. Nevertheless, there is a sudden increase in the average response time for the values less than $460$ ms of the update step. This means lowering the update step works out up to some point. By the values less than $460$ ms, the network gets fully congested and the negative impact of this congestion cancels the positive impact of more frequently updated news.

\begin{figure}
	\begin{center}
		\includegraphics[width=0.4\textwidth]{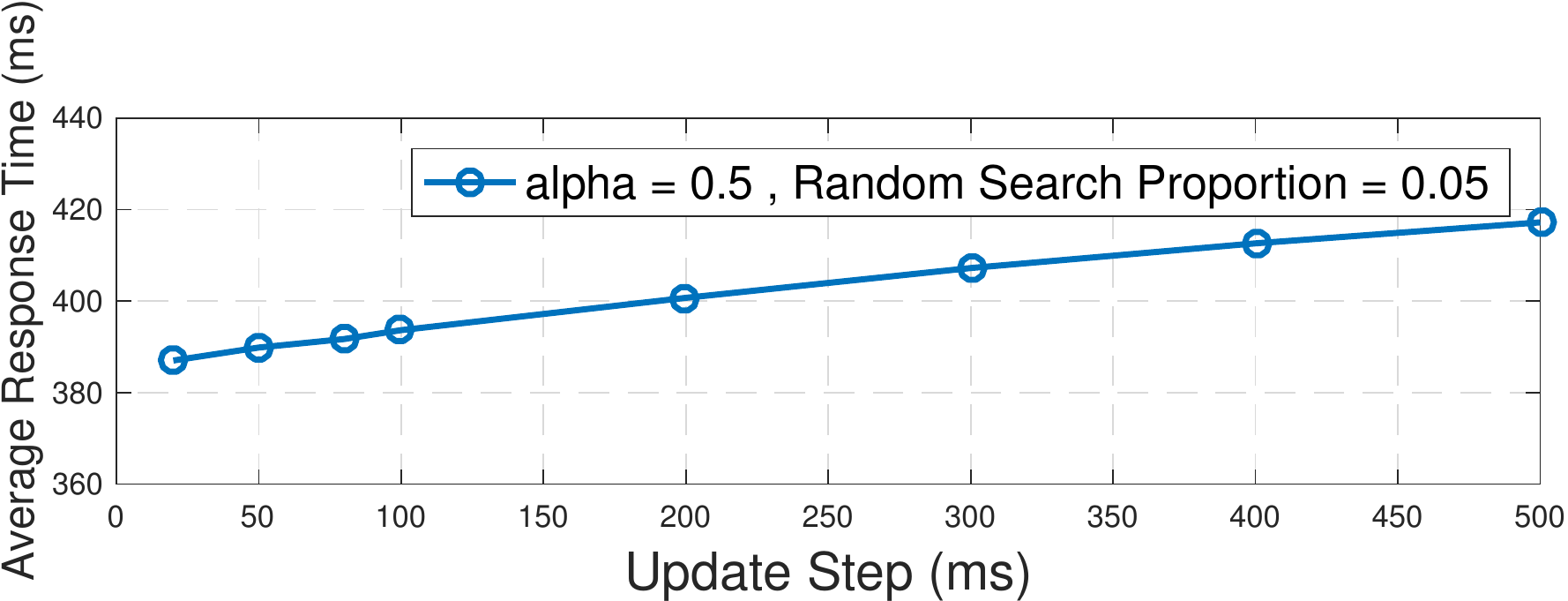}
	\end{center}
	\caption{The impact of update step of the proposed scheme on the average response time. The gradual slope reflects the uncritical impact of update step on the response time.}\label{fig:honeyPer}
\end{figure}

Similarly, Fig.~\ref{fig:honeyPer} investigates the update step of the proposed scheme vs. the average response time when the random search fraction is set to $0.05$. The gradual slope of the curve signifies the little influence of this parameter on the waiting time. This means while lowering the update step increases the congestion in the network, it does not drastically reduce the waiting time. So for further analysis, we set the update step to 500 ms. This is because it is close to the optimum update step of the periodic algorithm (460 ms) and also it causes less congestion.

\begin{figure}
	\begin{center}
		\includegraphics[width=0.4\textwidth]{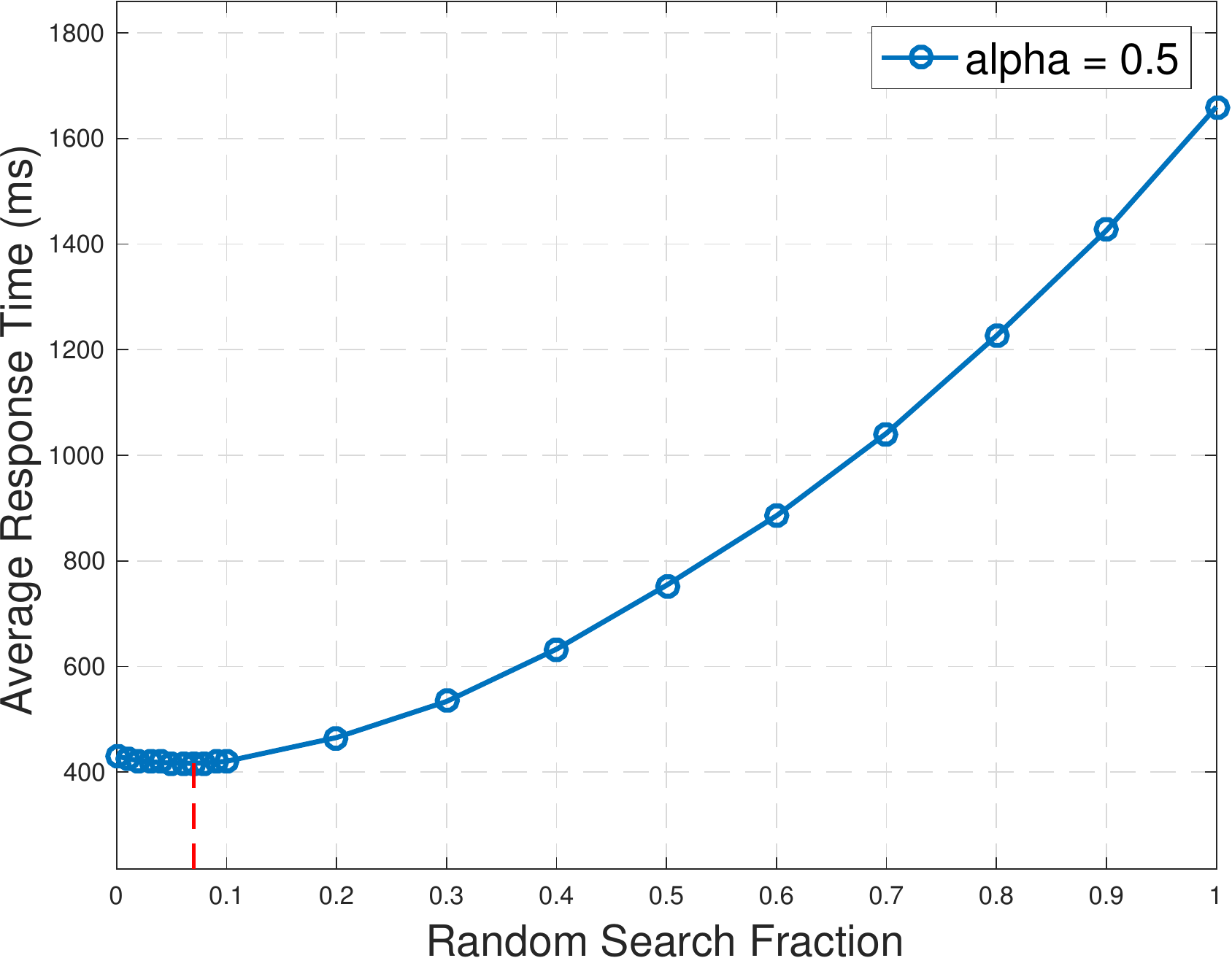}
	\end{center}
	\caption{Optimum random search fraction of the proposed scheme (in the interval $[0, 1]$). The update step for the proposed scheme is set to $500$ ms.}\label{fig:BigPic}
\end{figure}

\begin{figure}
	\begin{center}
		\includegraphics[width=0.4\textwidth]{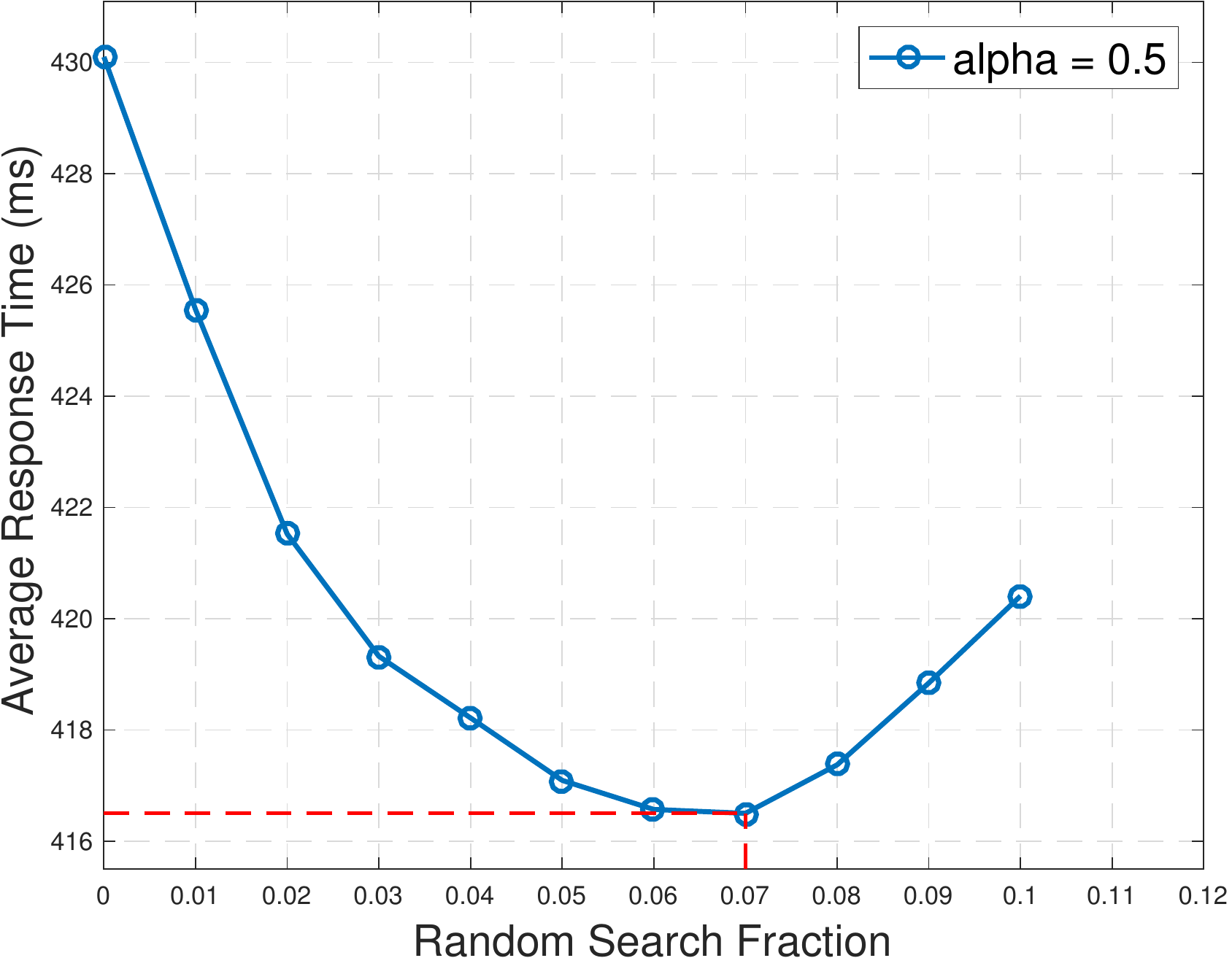}
	\end{center}
	\caption{The zoomed version of Fig.~\ref{fig:BigPic} on the interval $[0,12]$. The optimum random search fraction of the proposed scheme is $0.07$. The update step is set to $500$ ms.}\label{fig:Zoom}
\end{figure}

Next, we determine the optimum random search fraction in the proposed scheme. The least average response time is obtained when the random search factor is set to $0.07$, as shown in Fig.~\ref{fig:BigPic} and Fig.~\ref{fig:Zoom}, where the latter is the zoomed version of the former on the interval $[0 , 0.1]$.

\subsection{Results}
\begin{figure}
	\begin{center}
		\includegraphics[width=0.4\textwidth]{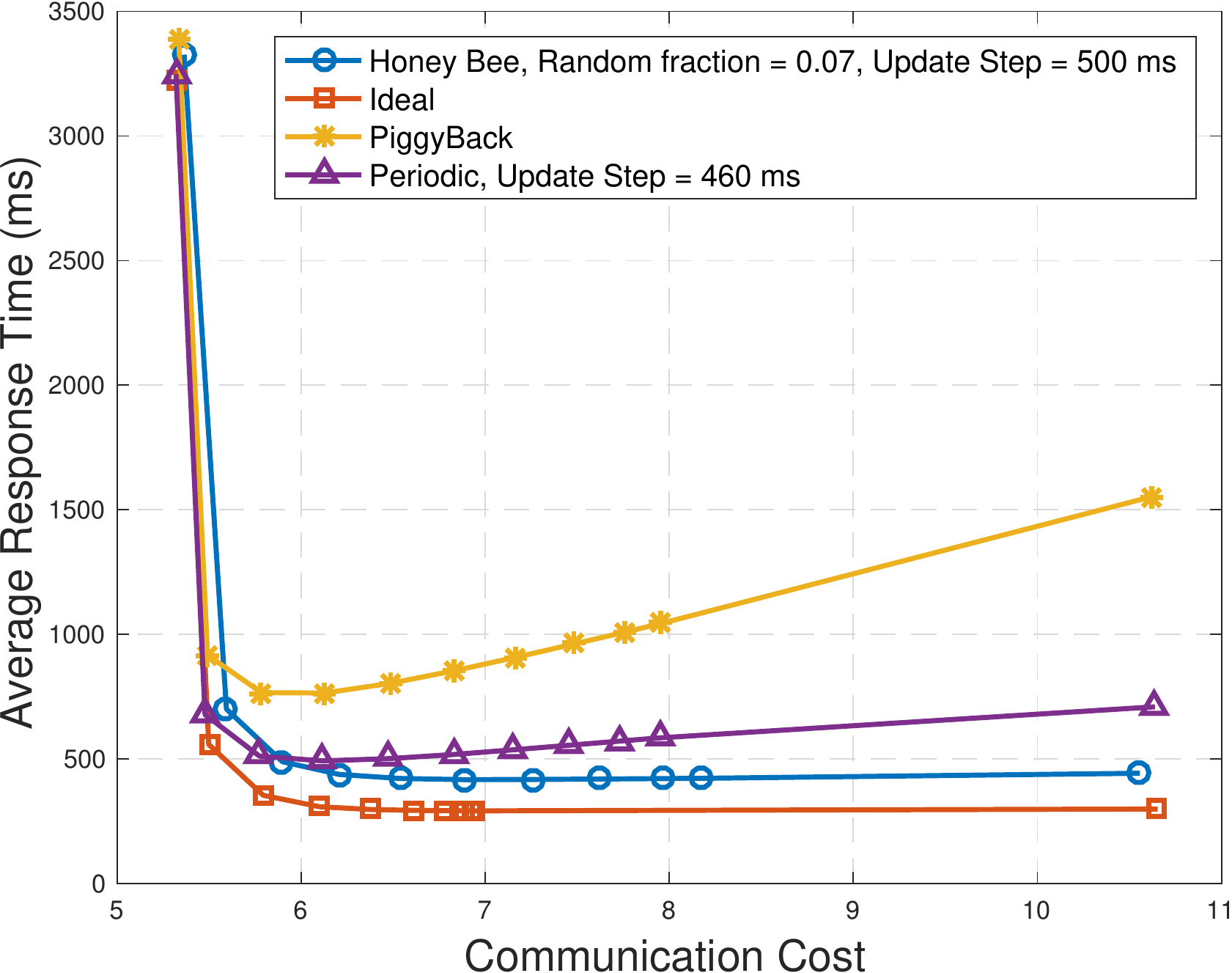}
	\end{center}
	\caption{Average response time, communication cost trade-off for the periodic, piggyback, the proposed, and ideal schemes.}\label{fig:TradeOff}
\end{figure}

Eventually, after determining the optimum values of the hyper-parameters, it is time to compare the proposed scheme with the existing ones. We will show how our proposed scheme surpasses the other schemes in managing the trade-off between communication cost and average response time.  Fig.~\ref{fig:TradeOff} illustrates the comparison of the schemes previously discussed: ideal-WMC, periodic-WMC, piggyback-WMC, and the proposed scheme. On each curve, there are eleven points, each corresponding to an $\alpha$ value for the WMC algorithm (Algorithm~\ref{alg:wmc}). At the leftmost point in each curve $\alpha=1$, where the communication cost is the only factor impacting WMC's decisions. As we go right, the $\alpha$ decreases by 0.1 at every midpoint to the rightmost point where the $\alpha = 0$. At the rightmost point, the only factor impacting WMC's decisions is servers' load. However at the midpoints, WMC decides based on a combination of load and cost as stated in Algorithm~\ref{alg:wmc}.

As can be seen in Fig.~\ref{fig:TradeOff}, the closest curve to the ideal scheme is our proposed scheme, and the trade-off is noticeably better managed in comparison to the other algorithms. The values of hyper-parameters are shown in the figure. Note that the proposed scheme excels the periodic despite the fact that its update step (500 ms) is more than periodic update step (460 ms). 

Another point to mention in Fig.~\ref{fig:TradeOff} is the positive slope of the periodic and piggyback curves at their rightmost parts. To explain this phenomenon, we should mention a previously explained note. On the one hand, as we go right on each curve, the load of the servers becomes a more important criterion to WMC algorithm (recall $\alpha$ varies from $1$ to $0$ from left to right). On the other hand, the news from other servers in piggyback and periodic schemes are less frequently updated and therefore less reliable compared to proposed and Ideal schemes. So, as we go right in the two curves, WMC's decisions become less reasonable because of their wrong information of the other servers. These wrong decisions cause an increase in the response time and therefore explains the positive slope of the curve. This effect is mitigated in the proposed scheme thanks to its random search feature, since it helps each server updates its knowledge of the other servers in a purposeful manner. Suppose a scenario in which a server's news of others shows another server is too much loaded. The periodic algorithm leaves it alone until the next period when the new information comes. However, the server may become desirable before the period ends.  The random search feature in the proposed scheme makes this update possible because it randomly learns news of other servers.

\begin{figure}
	\begin{center}
		\includegraphics[width=0.55\textwidth]{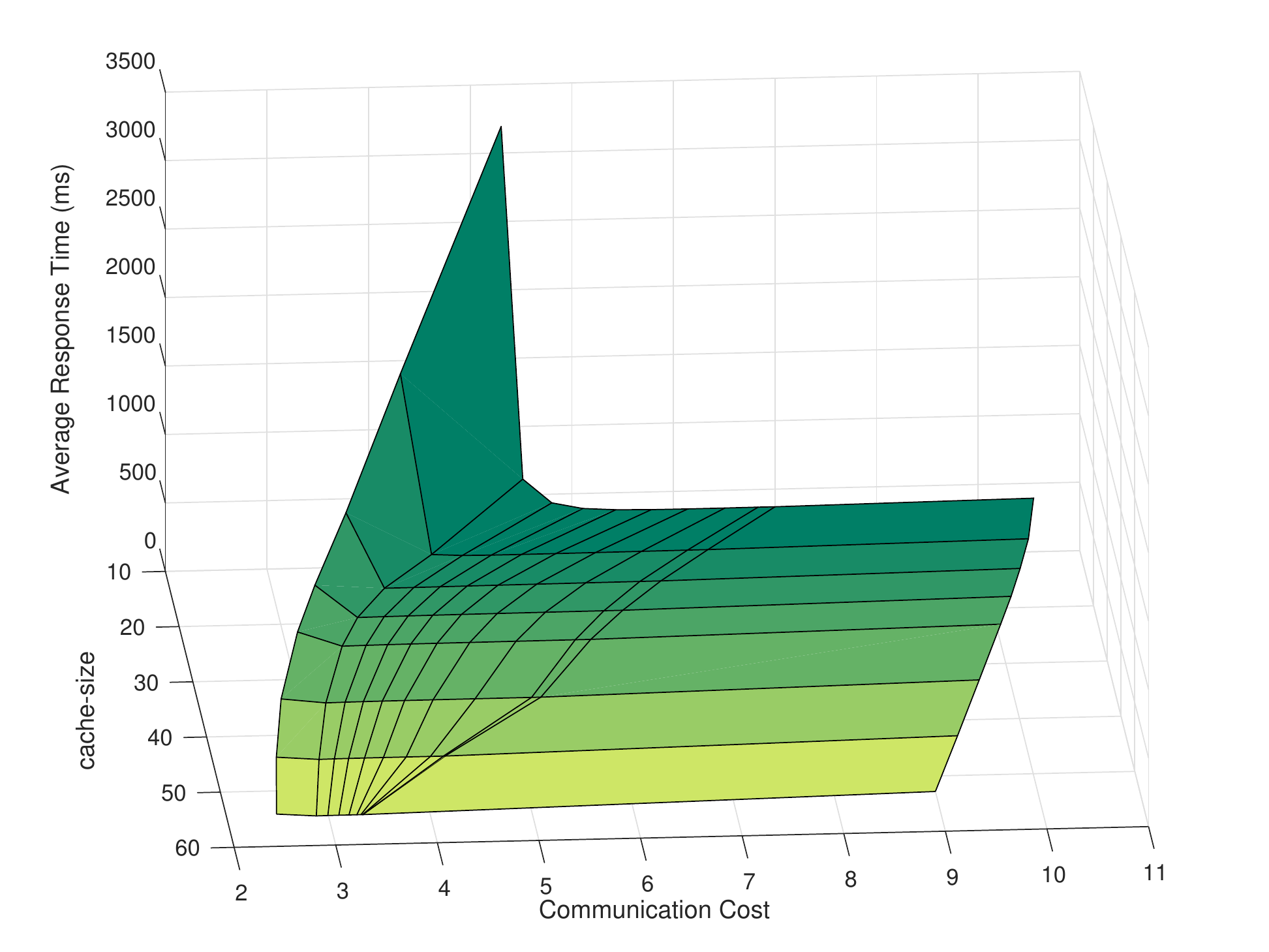}
	\end{center}
	\caption{The response time vs. communication cost of the proposed scheme for different cache sizes. For better view, please see the color version.}\label{fig:cachSize}
\end{figure}

It is also useful to show the impact of altering the cache size on the trade-off of the proposed scheme. In Fig.~\ref{fig:cachSize}, the cache size varies from $10$ to $60$ and the impact on the trade-off curved is shown. The back curve is what we saw in Fig.~\ref{fig:TradeOff} with the cache size $10$, and the front curve is the trade-off when the cache size is $60$. Expectedly, the more cache size allotted to each server, the less communication cost and response time is experienced. However, the overall shape of the trade-off is preserved.

%The extreme case, cache size 100, is the shortest, almost straight line owing to the fact that every server has all the files and there is no need to redirect the requests in most cases.

\section{Discussion and Concluding Remarks}\label{sec:Conclusion}
We investigated the trade-off between communication cost and average response time and proposed a meta-heuristic algorithm to achieve a better trade-off curve compared to the previous results. This scheme consists of two parts: an evaluation method (WMC), and an extension of the Honeybee scheme. This extension benefits from combining a layered version of the periodic update method and the piggyback. This combination leads to less network overhead while making the server's information about each other more accurate. 

To evaluate this scheme, we simulated it on our developed simulation tool, and the results reflect that the scheme manages this trade-off better than the other existing methods. Also, we have analyzed different servers' cache-size. Our results show the effect of cache size on the communication cost-average response time trade-off. However, the overall trend of the trade-off is preserved while cache size varies.
%and concluded that it is not of a significant influence on the trade-off curve. 

As a direction for future works, it would be more effective if we can unify a dynamic content placement strategy with such request redirecting schemes.

\section*{Acknowledgements}
	The authors would like to thank Pooya Shariatpanahi and Naeimeh Omidvar for their valuable discussions and feedback.

%Bibliography-------------------------------------------------------
\bibliographystyle{IEEEtran}
\bibliography{paper}

\end{document}